\documentstyle[prl,aps,multicol,epsf]{revtex}

\begin{document}
\draft

\title{Vortex Collisions: Crossing or Recombination?}

\author{Malek Bou-Diab$^{\,1}$, Matthew J.\ W.\ Dodgson$^{\,1,2}$,
and Gianni Blatter$^{\,1}$}

\address{$^{1\,}$Theoretische Physik, ETH-H\"onggerberg, CH-8093 Z\"urich, Switzerland}
\address{$^{2\,}$Theory of Condensed Matter Group, Cavendish Laboratory, Cambridge, CB3 0HE,
United Kingdom}

\date{\today}

\maketitle

\begin{abstract}
We investigate the collision of two vortex lines moving with viscous dynamics
and driven towards each other by an applied current. Using London theory in the
approach phase we observe a non-trivial vortex conformation producing
anti-parallel segments; their attractive interaction triggers a violent
collision. The collision region is analyzed using the time-dependent
Ginzburg-Landau equation. While we find vortices will always recombine through
exchange of segments, a crossing channel appears naturally through a double
collision process.
\end{abstract}

\pacs{PACS numbers: 74.60.Ec, 74.60.Ge}

\begin{multicols}{2}
\narrowtext

The important role that vortex collisions play in the non-equilibrium properties
of type-II superconductors was recognized long ago\cite{Josephson,Campbell}.
E.g., vortices subject to a longitudinal (parallel to field) current are
unstable towards helical expansion, leading to dissipation, but this instability
is contained if the vortices cannot cut through each other. This concept goes
back to Josephson\cite{Josephson}, who assumed that the flux--line cutting
process was energetically too costly to occur, a conclusion that has been 
rejected by Brandt {\it et al.} \cite{BrandtClemWalmsley} on the basis of
the first explicit calculation of the cutting energy. More recently, 
the extent to which entangled vortices can cut through each other has been
argued\cite{Marchetti} to determine the dynamics of the vortex--liquid phase,
which dominates much of the $H$-$T$ plane for high-$T_c$
superconductors\cite{Blatter}. Assuming an energy barrier $U_\times$ to vortex
crossing, an exponential dependence $\tau,\,\eta \propto \exp(-U_\times/T)$ of
the relaxation time $\tau$ and the viscosity $\eta$ of the vortex liquid was
predicted\cite{Obukhov,Cates}. Both flux transformer- and $c$-axis transport
experiments\cite{CFTE} are particularly suitable to test these ideas.

In the original calculation of the crossing energy $U_{\times}$, Brandt {\it et
al.}\cite{BrandtClemWalmsley} analyzed the intersection of two straight
vortices and calculated their configurational energy within London and
Ginzburg-Landau theory, see also Wagenleithner\cite{Wagenleithner}. Later,
Sudbo and Brandt\cite{Sudbo} found a lower barrier when all the conformal
degrees of freedom of an elastic line are included: the lines bend so as to take
advantage of the attraction between oppositely directed vortices. The energetics
of crossing in the presence of the surrounding vortex lattice, as well as
defects such as two-line and three-line twists, have been calculated, both in
the London model\cite{Schonenberger} and using the lowest Landau level
approximation\cite{Dodgson}. While these previous studies have concentrated on
metastable configurations, here we are concerned with the vortex scattering
process itself and particularly with the output topology after the collision:
vortices, being line objects, can end up in two alternative topological
configurations, reswitched or crossed (see Fig.\ 1).
In this letter we analyze the conditions selecting between these two scattering
channels.

Vortex collisions have attracted a lot of interest in a variety of fields,
ranging from turbulence in superfluid Helium\cite{Schwarz} to galaxy formation
via cosmic strings\cite{Shellard}. While the vortex dynamics is hamiltonian
(Schr\"odinger type or massive) in these systems, the vortex dynamics in type II
superconductors is dissipative in general, implying that a collision has to be
driven by an external current ${\bf J}$. Also, while the usual scattering of
particles is characterized by a small number of parameters, the vortex
collisions studied below involve line objects with an infinite number of degrees
of freedom that eventually could influence the result of the collision. Here
we study vortex collisions for a set of typical initial geometries and
drives and extract the generic information at the end.
\begin{figure}
\centerline{\epsfxsize = 6.0cm \epsfbox{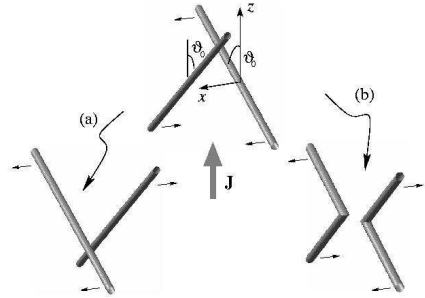}} \narrowtext\hspace{0.5cm}
\caption{Initial configuration of two colliding vortices enclosing an angle
$2\vartheta_0$  and pushed towards each other along the $x$-axis by the
current density ${\bf J}$ directed parallel to the $z-$axis.
The collision terminates in one of two topologically different configurations:
(a) the vortices cut through each other; (b) the vortices recombine through
exchange of segments.} \label{fig:1}
\end{figure}
\vspace{-0.3cm}
Below, we study the collision on two different length scales: at large
distances, of the order of the penetration depth $\lambda$, the vortices behave
as elastic lines interacting through currents and a London description is valid.
On the other hand, as the vortices approach on distances of order the coherence
length $\xi$, the topology of the lines around the collision point is not well
defined within the London scheme and we have to adopt the time-dependent
Ginzburg-Landau formalism\cite{Schmid} (TDGL). The vortices then are described
by a complex wave function which accounts for the evolution of the vortex core
structure. Here, we consider the situation where these two scales are well
separated, i.e., superconductors with a large Ginzburg-Landau parameter $\kappa
= \lambda/\xi \sim 100$ typical for high temperature superconductors.

Our results in the {\it London regime} show that even for nearly parallel
vortices and weak driving currents the two vortices always collide: the drag
force acting on the vortex ends always over-compensates the local repulsive
force. In addition, the induced torque twists the initially repulsive vortex
segments towards a locally anti-parallel configuration with an attractive
interaction, resulting in a violent collision. From our {\it TDGL analysis} at
small scales we conclude that the vortices always recombine when the cores
collide, independent of drive and of the angle of collision. Nevertheless, a
crossing channel can appear in the system through a double collision mechanism:
we find that for our geometry the vortex recombination naturally leads to an
intermediate configuration which is not `free', always enforcing a second
collision. This second collision restores the original topology of the
individual lines and results in a crossing configuration which is 
asymptotically free. These findings then lend support to an early 
proposal by Clem \cite{Clem81} for a similar double-cutting scenario for 
colliding vortex lines. 

The initial configurations studied here are shown in Fig.\ 1: the
two straight vortex lines ${\cal V}_{1}$ and ${\cal V}_{2}$ enclose an
angle $2 \vartheta_0$ and are driven towards each other by an external
current density ${\bf J}$. For large distances $d > \xi$ between the
colliding vortices the system can be described within the London
approximation\cite{Tinkham,Blatter}. The vortices are line objects
parametrized by the positions ${\bf r}_{\mu}(z)$, where $\mu=1,2$ for
the vortices ${\cal V}_{1}$ and ${\cal V}_{2}$. We concentrate on isotropic
superconductors (generalization to the anisotropic case is done by
rescaling\cite{BGL}) and write the free energy functional in the form
\begin{eqnarray}
\label{eq:London}
  {\cal F}_L[\{ \hbox{\bf r}_\mu \}]
  &=& \frac{\varepsilon_0}{2} \sum_{\mu,\nu=1}^{2}
  \int d\hbox{\bf r}_\mu \cdot d\hbox{\bf r}_\nu
  \frac{ e^{-\sqrt{|\hbox{\bf r}_\mu-\hbox{\bf r}_\nu|^{2}
  +\xi^{2}}/\lambda}}{\sqrt{|\hbox{\bf r}_\mu-\hbox{\bf r}_\nu |^{2}
  +\xi^{2}}} \nonumber \\
  &+& \varepsilon_0 c_{0}\sum_{\mu=1}^{2}\int | d{\bf r}_\mu |;
\end{eqnarray}
the first term describes the pairwise interaction between vortex segments, with
$\varepsilon_{0}=(\Phi_0/4\pi\lambda)^{2}$ the basic energy scale
($\Phi_0=hc/2e$ is the flux unit). The terms $\mu=\nu$ correspond to the self
energy of the vortices while the terms $\mu\neq\nu$ describe the interaction
between them. The second part in (\ref{eq:London}) accounts for the energy of
the vortex cores; the value $c_{0}=0.38$ is found from comparing the vortex
energy within the London theory with the value derived from a Ginzburg-Landau
analysis\cite{Hu}.

Below we concentrate on the vortex ${\cal V}_{1}$; the configuration of vortex
${\cal V}_{2}$ follows from symmetry. The forces acting on each point of
${\cal V}_{1}$ are the Lorentz force due to the current density ${\bf J}$, the
friction force generated by the dissipation in the vortex core, and the elastic
and interaction forces as given by the functional derivative of the free energy
functional (\ref{eq:London}). The dimensionless equation of motion for a point
${\bf r}_{1}$ on vortex ${\cal V}_{1}$ then takes the form
\begin{eqnarray}
  \frac{\partial{\bf r}_{1}}{\partial t}
  =&-&\sum_{\mu=1}^{2}\int_{{\cal V}_{\mu}}(1+\rho_{\mu}^{-1})
  \frac{\exp{(-\rho_{\mu})}}{\rho_{\mu}^{2}}
  {\bf n}_{1} \wedge ({\bf dr}_{\mu} \wedge {\bf r}_{1\mu}) \nonumber \\
   &-&c_{0} [{\bf n}_{1} \wedge ({\bf n}_{1}\wedge {\bf k}_{1})]
    +\kappa ({\bf J} \wedge {\bf n}_{1}), \label{eq:motion}
\end{eqnarray}
where ${\bf r}_{1\mu}={\bf r}_{1}-{\bf r}_{\mu}$ and $\rho_{\mu}=\sqrt{|{\bf
r}_{1\mu}|^{2} + 1/\kappa^{2}}$; ${\bf n}_{1}={\bf r}_1'(z)/|{\bf r}_1'(z)|$
denotes the tangent vector and ${\bf k}_{1}={\bf r}_1''(z)/|{\bf r}_1'(z)|^{2}$
($'=d/dz$ and $''=d^{2}/dz^{2}$). Note that Eq.\ (\ref{eq:motion}) is
independent of the parameterization of the line; the forces acting on the
segments are locally orthogonal. We have chosen units: $[r]=\lambda$,
$[Force]=\varepsilon_{0}$, $[J] = j_0 = \varepsilon_{0} c / \xi \phi_{0}$ with
$j_0$ of the order of the critical current density, and $[t] = \eta_{l}\lambda^{2}/
\varepsilon_{0}$ with $\eta_{l} = \phi_{0}^{2} / 2\pi\xi^{2}c^{2}\rho_{n}$ the
viscosity per unit length as given by Bardeen and Stephen\cite{Tinkham}
($\rho_{n}$ denotes the normal state resistivity). Initially, the two vortex
lines are separated by a distance $d \approx \lambda$ and enclose an angle $2
\vartheta_{0}$, see Fig.\ 1. In order to describe the vortex collision including
all dynamical and configurational degrees of freedom we solve the equation of
motion numerically using a second-order Runge-Kutta method. The elastic
forces help stabilizing the algorithm provided that the distance $\delta$ 
between neighboring points remains smaller than $1/\kappa$ and the time 
step is $\sim \delta^2$. 

The integration of the equation of motion for various angles $\vartheta_{0}$ and
current densities ${\bf J}$ reveals two generic types of collisions depending on
the initial angle. We first discuss the results for a `steep' collision with a
small initial angle $\vartheta_{0} < \pi/4$, see Fig.\ 2. As the vortices enter
the range of interaction, the vortex segments close to the $x$-axis are subject
to a strong repulsive interaction. This repulsive force is compensated by the
Lorentz force due to the applied current density $J$ and the drag force
generated by the vortex ends. The latter is the result of the elastic forces in
the line: the vortex ends far away from the center do not interact and are
pushed by the Lorentz force towards the origin, their motion dragging the
central part along towards collision. With decreasing separation the torque
generated by the interaction grows and twists the central parts of the lines,
thereby increasing the angle between the central segments. As this angle
increases beyond $\pi/2$ the interaction turns attractive, the central vortex
segments turn anti-parallel and collide. After the collision these segments
remain bound at a distance $d \lesssim \xi$ with both cores overlapping: the
Lorentz force due to $J$ is not sufficient to separate the anti-parallel
segments until the drag provided by the vortex ends has become large. Note that
even a small Lorentz force (e.g., for a small angle $\vartheta_{0}$ and current
drive $J$) is sufficient to trigger a collision due to the line twist. In the
second, `flat' type of collision with $\vartheta_{0} > \pi/4$, all forces drive
the vortices toward each other and they collide rapidly; the collision scenario
then simplifies without the initial repulsive phase, the central vortex segments
twisting smoothly and colliding violently.
\begin{figure}
\centerline{\epsfxsize=7.5cm\epsfbox{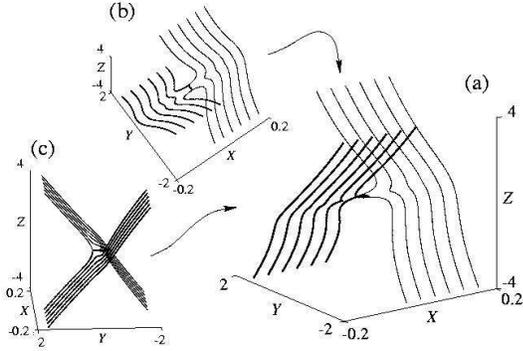}} \vspace{0.3cm} \caption{Vortex
configurations at constant time steps for a `steep' collision with initial
angle $\vartheta_{0}=\pi/8$; a driving current $J=0.4 J_{0}$ has been chosen.
Note the various behaviors of the central segments during the collision:
repulsion, twist, attraction, and collision with binding ((a) side view, (b)
top view, (c) front view, units in $\lambda$).}
\label{fig:2}
\end{figure}
\vspace{-0.2cm}

At distances $\sim \xi$ the London model is not valid; the vortex
cores overlap and the interaction is dominated by the condensation rather than
electromagnetic energies. We then refine the discussion of the vortex collision
by solving the time-dependent Ginzburg-Landau equations\cite{Schmid} within a
small box of size $\sim 10 \xi$ around the collision point. The vortex
configuration is described by a complex wave function $\psi({\bf r},t)$
undergoing a time evolution
\begin{equation}
\label{eq:tdgl}
  (\partial_t-i\tilde{\phi})\psi
  =\left({\bf \nabla}/\kappa + i{\bf A}\right)^{2}\psi
   + (1 - |\psi|^{2})\psi,
\end{equation}
where $\tilde{\phi}$ is the electrochemical potential and we use the units:
$[r]=\lambda$, $[{\bf A}]=\sqrt{2}\lambda H_{c}$,
$[\psi]=\sqrt{|\alpha|/\beta}$, $[t]=\tau$ $=\hbar\gamma/|\alpha|$, and
$[\tilde{\phi}]=\hbar/2e\tau$. Here, $\alpha$ and $\beta$ are parameters in the
Ginzburg-Landau free-energy, $H_{c}$ is the thermodynamic critical field, and
$\gamma$ is the damping coefficient, see \cite{Schmid}. For a clean
superconductor $\gamma=1/2 k_{\rm\scriptscriptstyle F} \xi_{0}$ ($\xi_{0}$ is
the BCS coherence length) resulting in a ballistic time scale
$\tau=\xi_{0}/v_{\rm\scriptscriptstyle F}$, while in the dirty case $\gamma =
3/2 k_{\rm\scriptscriptstyle F}l$ ($l$ the mean free path) with a diffusive time
$\tau =3 \xi_{0}^{2}/l v_{\rm\scriptscriptstyle F}$ ($k_{\rm\scriptscriptstyle
F}$ and $v_{\rm\scriptscriptstyle F}$ denote the Fermi wave vector and Fermi
velocity). Eq.\ (\ref{eq:tdgl}) is completed by the Maxwell equations and the
constitutive relations for the superconductor\cite{Schmid}, of which we need the
expressions ${\bf j}_n = -2\Sigma(\partial_t {\bf A} + \nabla {\tilde
\phi}/\kappa)$ and ${\bf j}_s = i[\psi^{*}({\bf \nabla}/\kappa + i {\bf A})\psi
- {\rm c.c.}]$ for the normal and superconducting current densities with $\Sigma
= m\beta/4e^{2}\hbar\rho_n\gamma$; the continuity equation $\partial_t \rho +
\nabla\cdot({\bf j}_n + {\bf j}_s) = 0$ with $\rho$ the charge density can then
be used to determine the potential $\tilde{\phi}$. Using the values of $\beta$
and $\gamma$ derived from microscopic theory\cite{Schmid} the coefficient
$\Sigma$ takes the value $\Sigma=14\zeta(3)/\pi^{4}$ in the dirty case, while
$\Sigma=4l/\pi^{3}e^{-C}\xi_{0}$ for a clean superconductor ($C$ is the Euler
constant); below we concentrate on the parameter-free dirty situation. As we are
interested in the core region (i.e., scales smaller than $\lambda$) we may
ignore transverse screening by choosing a gauge with ${\bf A} = 0$. Furthermore,
neglecting weak charging effects\cite{Schmid} the requirement of divergence free
flow,
\begin{equation}
\label{eq:stdgl}
   \Sigma{\bf \nabla}^{2}\tilde{\phi}
  =(i/2)[\psi^{*}{\bf \nabla}^{2}\psi-\psi{\bf \nabla}^{2}\psi^{*}],
\end{equation}
determines the electrochemical potential $\tilde\phi$ needed to push the normal
current density ${\bf j}_n$ through the vortex core. We use the initial
condition
\begin{equation}
\label{eq:initial} \psi({\bf
r},t=0)=\psi_{1}\psi_{2}\sqrt{1-\nu^{2}}\exp(-i\kappa\nu z),
\end{equation}
where $\psi_{i},~i=1,\,2$ denote the wave functions for the vortices ${\cal
V}_{i}$ [$\psi(r,\varphi)=(\kappa r/\sqrt{\kappa^2 r^2+2})\exp{i\varphi}$]
appropriately rotated and translated. The last factor in (\ref{eq:initial})
describes the driving current $j_{s}=2\nu(1-\nu^{2})$ along the $z-$axis with a
tuning parameter $\nu$ and a prefactor that takes into account the reduction in
the superfluid density due to the applied current. This applied current is
maintained during the time evolution through the boundary conditions
${\partial}_{z}\psi=-i\kappa \nu\psi$ on the top and bottom of the box. We
always adapt the dimensions of the box such that the cores penetrate through the
sides in order not to disturb the applied current density; we use two types of
boundary conditions on the box sides to make sure that our results are not
influenced by our specific choice: {\it i)} the boundary condition ${\bf
\nabla}_{\bot} \psi=0$ guarantees that no currents leave the box through the
sides, and {\it ii)} we impose the boundary condition $\psi({\bf r},t) =
\psi_{1}({\bf r}-{\bf v}t)\psi_{2}({\bf r}+{\bf v}t)\sqrt{1-\nu^{2}}
\exp(-i\kappa\nu z)$ (c.f.\ (\ref{eq:initial})) with the velocity ${\bf v}$
adapted to the local velocity of the vortex close to the boundary. We start with
the vortices separated a distance of a few $\xi$ in order to have the wave
functions relax before they enter the collision region and integrate Eq.\
(\ref{eq:tdgl}) using the Forward-Euler method which converges well for our
dissipative dynamics. The Poisson equation (\ref{eq:stdgl}) for $\tilde{\Phi}
({\bf r},t)$ is integrated at each time step using a conjugate gradient
method\cite{comment}.

A systematic survey for different angles and applied current densities smaller
than the depairing current provides the following results: any collision,
independent of initial angle or drive, makes the vortices recombine through
exchange of segments, see snapshots (a) to (c) in Fig.\ 3. The resulting
reswitched configuration is not `asymptotically free': the drive current now
acts differently on the newly built vortex lines and pushes the two vortex ends
of each line into opposite directions, inducing a twist of the central vortex
segments. As these segments turn around, the Lorentz force changes direction and
the vortices are driven towards a new collision (snapshot (d)). The second
collision (snapshot (e)) then enforces another recombination, resulting in a
`crossing' topology as if the original colliding vortices had cut through each
other (snapshot (f); we have found this double collision to occur in all our
simulations with $\vartheta_0/\pi \in [1/12,5/12]$ and $j/j_0 \in [0.15,0.75]$). We
thus find a rather unexpected result: while each (local) collision induces a
change in topology through recombination, double collisions occur naturally due
to geometrical constraints and lead to vortex crossing --- the asymptotic regime
where the vortices can separate is only reached when the vortices have crossed
after a second collision.
\begin{figure}
\centerline{\epsfxsize=6.9cm\epsfbox{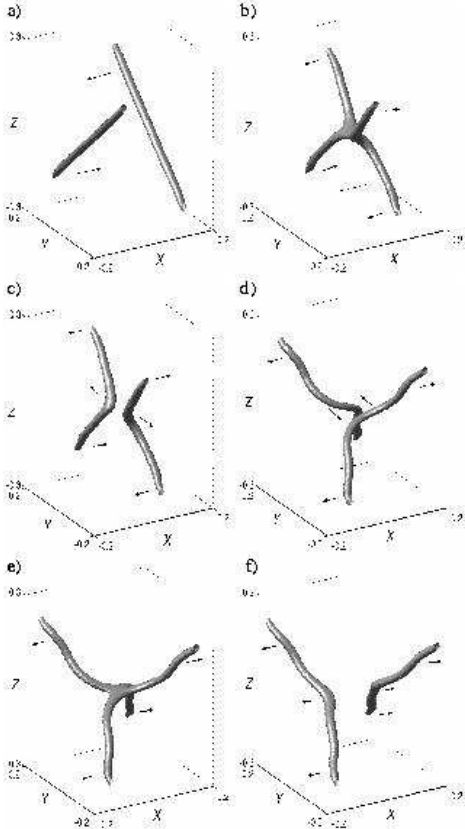}} \vspace{0.3cm} \caption{Time
evolution of a vortex collision calculated within TDGL theory for
$\vartheta_{0}=\pi/4$ and $J=0.6J_0$. The time sequence (from left to right and
top to bottom) reveals a double collision process: (a)--(c) first collision with
switched segments due to recombination; (d) central segments twist and prepare
for second collision (e); (f) second recombination with asymptotically free
vortices in the crossing topology.} \label{fig:3}
\end{figure}
\vspace{-0.2cm}
While we have studied the generic situation of two individual colliding
vortices, vortex collisions occur and are important in other situations, e.g.,
within the vortex solid or liquid phase of (high-$T_c$) superconductors. The
`asymptotic' regimes defining the initial and final states of the collision may
then look quite different and the second collision producing vortex crossing,
which is generic to the above geometry, may be absent. An example for such a
situation is the expansion of a vortex twist\cite{Schonenberger} in a vortex
lattice or liquid phase. The interaction with neighboring vortices inhibits the
second twist of the segments and each collision terminates after one encounter
--- as a result, the vortex twist (or ring) smoothly expands through
recombination processes. Such recombination processes have also been suggested
to relax the stress in driven vortex systems\cite{Cates}. Furthermore, vortex
recombination has been assumed to help relax the pitch of spiral vortices in
current carrying superconducting cylinders subjected to a longitudinal
field\cite{Clem}.

Summarizing, colliding vortices in type-II superconductors always recombine
through exchange of segments. However, the line nature of the vortices can
enforce a second collision, opening a vortex crossing channel through a double
collision. The two cases are easily distinguished through careful inspection of
the asymptotic constraints enforced on the collision.

We thank V. Geshkenbein, M. Troyer, and M. Cates for stimulating discussions, and
the Swiss National Foundation for financial support.

\end{multicols}

\end{document}